\documentclass[conference,10pt]{IEEEtran}
\usepackage{graphicx}
\usepackage{cite,amssymb,amsmath,bm}
\usepackage{mathrsfs}
\usepackage[pagewise]{lineno}
\usepackage{color,soul}
\usepackage{tabulary}
\usepackage{multirow,multicol}
\usepackage{cases}
\usepackage{stfloats}
\usepackage{url}
\usepackage{booktabs}
\usepackage{algpseudocode}
\usepackage{algorithm,algpseudocode}
\usepackage{algorithmicx}
\usepackage[normalem]{ulem}

\usepackage{tikz}

\ifCLASSOPTIONcompsoc
    \usepackage[caption=false, font=normalsize, labelfont=sf, textfont=sf, subrefformat=parens, labelformat=parens]{subfig}
\else
\usepackage[caption=false, font=footnotesize, subrefformat=parens, labelformat=parens]{subfig}
\fi

\usepackage{amsmath,amssymb}
\usepackage{mathtools} 
\usepackage{xspace}

\newcommand{\Figref}[1]{Fig.~\ref{#1}}

\newcommand{\herm}{^\text{H}}

\newcommand{\trans}{^\text{T}}

\newcommand{\bZ}{\mathbf{Z}}

\newcommand{\bh}{\mathbf{h}}

\newcommand{\bbf}{\mathbf{f}}

\newcommand{\bI}{\mathbf{I}}

\newcommand{\bR}{\mathbf{R}}

\newcommand{\bC}{\mathbf{C}}
\newcommand{\bD}{\mathbf{D}}

\newcommand{\bb}{\mathbf{b}}
\newcommand{\bc}{\mathbf{c}}
\newcommand{\be}{\mathbf{e}}

\newcommand{\bv}{\mathbf{v}}

\newcommand{\bp}{\mathbf{p}}

\newcommand{\bzero}{\boldsymbol{0}}

\newcommand{\CN}{\mathcal{CN}}
\newcommand{\tc}{\tau_{\textsc{c}}}

\newcommand{\boundellipse}[3]
{(#1) ellipse [x radius=#2,y radius=#3]
}

\DeclareMathOperator{\tr}{tr}

\newcommand{\EX}[1]{\mathsf{E}\left\{{#1}\right\}}

\newcommand{\C}{\mathbb{C}}

\newcommand{\tauc}{\tau_\mathrm{c}}

\newcommand{\norm}[1]{{ \left\Vert #1 \right\Vert }}

\newcommand{\tp}{\tau_{\mathrm{p}}}
\newcommand{\td}{\tau_{\mathrm{d}}}
\newcommand{\tu}{\tau_{\mathrm{u}}}

\makeatletter
\def\@setsize#1#2#3#4{
    \@nomath#1
    \let\@currsize#1
    \baselineskip #2
    \baselineskip \baselinestretch\baselineskip
    \parskip \baselinestretch\parskip
    \setbox\strutbox \hbox{
        \vrule height.7\baselineskip
            depth.3\baselineskip
            width\z@}
    \skip\footins \baselinestretch\skip\footins
    \normalbaselineskip\baselineskip#3#4}
\makeatother

\makeatletter
\newcommand{\setstretch}[1]{
    \def\baselinestretch{#1}%
    \@currsize
    }
\makeatother



\def\BibTeX{{\rm B\kern-.05em{\sc i\kern-.025em b}\kern-.08em
    T\kern-.1667em\lower.7ex\hbox{E}\kern-.125emX}}

\algnewcommand{\IfThen}[2]{
  \State \algorithmicif\ #1\ \algorithmicthen\ #2}

\newcommand{\comment}[1]{}

\newcommand{\csubfloat}[2][]{%
  \makebox[0pt]{\subfloat{#2}}%
}
\newcommand{\centerhfill}[1][\quad]{\hspace{\stretch{0.5}}#1\hspace{\stretch{0.5}}}

\begin{document}
\begin{figure*}[t!]
\normalsize
Paper accepted for presentation in IEEE ICC 2022 - IEEE International Conference on Communications. 

\

\textcopyright~2022 IEEE. Personal use of this material is permitted.  Permission from IEEE must be obtained for all other uses, in any current or future media, including reprinting/republishing this material for advertising or promotional purposes, creating new collective works, for resale or redistribution to servers or lists, or reuse of any copyrighted component of this work in other works.
\vspace{17cm}
\end{figure*}

\title{The Promising Marriage of Mobile Edge Computing and Cell-Free Massive MIMO}
\author{
\IEEEauthorblockN{Giovanni Interdonato and Stefano Buzzi}
\IEEEauthorblockA{Dept. of Electrical and Information Engineering, University of Cassino and Southern Latium, Cassino, Italy \\
\{giovanni.interdonato, buzzi\}@unicas.it}}

\maketitle

\begin{abstract}
This paper considers a mobile edge computing-enabled cell-free massive MIMO wireless network. An optimization problem for the joint allocation of uplink powers and remote computational resources is formulated, aimed at minimizing the total uplink power consumption under latency constraints, while simultaneously also maximizing the minimum SE throughout the network.
Since the considered problem is non-convex, an iterative algorithm based on sequential convex programming is devised.
A detailed performance comparison between the proposed distributed architecture and its co-located counterpart, based on a multi-cell massive MIMO deployment, is provided. Numerical results reveal the natural suitability of cell-free massive MIMO in supporting computation-offloading applications, with benefits over users' transmit power and energy consumption, the offloading latency experienced, and the total amount of allocated remote computational resources.
\end{abstract}


\vspace*{-2mm}
\section{Introduction}
Cooperative distributed massive multiple-input multiple-output (MIMO) systems, also known as \textit{cell-free} massive MIMO systems~\cite{Ngo2017b,Interdonato2019,Buzzi2019c,cellfreebook}, are envisioned to be a key-enabling technology for \textit{beyond}-5G mobile communication systems.
Cell-free massive MIMO is theoretically able to provide unprecedented levels of data rates and ubiquitous connectivity by leveraging an extraordinary macro-diversity gain, inter-cell interference mitigation, and user proximity.

Over the past few years, we have witnessed to an exponential growth of computation-intensive applications with strict latency requirements for mobile devices, e.g., online gaming and augmented reality. An approach to indirectly increase the computing capabilities of the devices, and prolongs their battery lifetime is to delegate their computational tasks to the network, specifically to \textit{network edge servers}. This approach is known as \textit{mobile edge computing} (MEC) or mobile-edge computation offloading~\cite{Yang2008,Mach2017,Mao2017}.

Most works on MEC optimize the interplay between the amount of computational tasks to offload, the latency due to the offloading operation, and the energy consumption of the mobile devices. Some works minimize the energy consumption under delay constraints~\cite{You2017,Zeng2020b}, or the remote computational time under energy consumption constraints~\cite{Ren2018}. 
With the advent of MIMO, and especially massive MIMO technology, research works on MEC focus on capitalizing the higher spectral and energy efficiencies, provided by the joint coherent transmission/reception from multiple antennas, to improve the offloading efficiency~\cite{Sardellitti2015,Nguyen2019}.
In~\cite{Pradhan2020}, the authors firstly utilize a successive inner convexification framework to minimize the total transmit power of the devices under latency constraints, then compare this solution with a low-complexity supervised deep learning approach. 
In~\cite{Hao2019}, the authors proposes a low-complexity algorithm, based on alternating optimization, to jointly optimize uplink powers and computing resources for a massive MIMO-enabled heterogeneous network with MEC, under the assumption of maximum-ratio combining (MRC) detector.
The main common conclusion of these works is that the offloading efficiency grows with the number of antennas of the massive MIMO base station.
The performance of cell-free massive MIMO with MEC functionalities has been recently explored in~\cite{Mukherjee2020}, assuming that a user can request computation offloading either to the central MEC server or to one of its serving APs. Hence, such a system implements a cell-free user-centric approach from the radio resource allocation viewpoint, but it does not from the MEC perspective. Following on this research track, we explore the potential benefits of jointly optimizing radio and computational resources in a cell-free massive MIMO system with user-centric MEC approach.
 
\textbf{Contribution.} We propose a MEC-enabled cell-free massive MIMO architecture implementing, unlike prior studies, a user-centric approach both from the radio and the computational resource allocation perspective. In particular, we formulate an optimization problem for jointly allocating users' transmit powers and the remote computational resources for offloading. Similarly to~\cite{Sardellitti2015,Hao2019}, we minimize the total uplink power under latency constraints, but unlike prior studies, we simultaneously maximize the minimum SE throughout the network.
We devise a successive convex approximation (SCA) algorithm for efficiently solving the optimization problem.
Compared to that in~\cite{Hao2019}, it additionally includes the user-centric clustering framework, a distributed allocation of the computational resources, and its formulation is generalized for any combining scheme.
Finally, we provide a detailed performance comparison between the proposed distributed architecture and its centralized counterpart, showing the natural suitability of cell-free massive MIMO in supporting computation-offloading applications. 

\section{System Model} \label{sec:sysmodel}

We consider a CF-mMIMO system operating in TDD mode, and consisting of $L$ geographically distributed access points (APs), equipped with $M$ antennas each, that are connected through a fronthaul network to a central processing unit (CPU). The APs coherently serve $K$ single-antenna users in the same time-frequency resources.      
The conventional block-fading channel model is considered. Each coherence block has length $\tauc$ samples and accommodates pilot-based uplink training, uplink and downlink data transmission, such that $\tauc = \tp + \tu + \td$, where $\tp$, $\tu$ and $\td$ are respectively the training duration, the uplink and the downlink data transmission duration, respectively.  
The system model resembles that described in~\cite[Section 4.1--4.2]{cellfreebook}, hence we limit its description in this section due to space constraints.  
The channel between the $k$-th user and the $l$-th AP is denoted by $\bh_{lk} \in\! \C^{M}$, with $\bh_{lk}\! \sim\! \CN(\bzero, \bR_{lk})$, and $\bR_{lk}\! \in\! \C^{M\!\times\!M}$ being the spatial correlation matrix. The corresponding large-scale fading coefficient is defined as $\beta_{lk} = \tr(\bR_{lk})/M$. 
The channel between the $k$-th user and all APs in the system is obtained by stacking the channel vectors $\bh_{lk}$, $\forall l$ as $\bh_k\! =\! [\bh_{1k}\trans \cdots \bh_{Lk}\trans]\trans \in \C^{LM}$. The channel vectors of different APs are reasonably assumed to be independently distributed. As a consequence, we have $\bh_{k}\! \sim\! \CN(\bzero, \bR_{k})$, where $\bR_{k}\! =\! \text{diag}(\bR_{1k}, \ldots, \bR_{Lk}) \in \C^{LM\! \times\! LM}$ is the block-diagonal spatial correlation matrix related to user $k$.   

\medskip

\textbf{Centralized Uplink Training.} 
During the uplink training all the $K$ users synchronously send a pre-determined pilot sequence (of length $\tp$) drawn by a set of $\tp$ orthonormal vectors. 
We assume that channel estimation is performed by the CPU according to the linear minimum-mean square error (MMSE) estimation criterion.
We denote the MMSE estimate of the channel between the $k$-th user and the $l$-th AP as $\hat{\bh}_{lk}$. The channel estimation error $\tilde{\bh}_{k}$ is independent of the estimate, and its covariance matrix is denoted by $\bC_{lk}$.
All the estimates for a specific user $k$ are collected in the vector $\hat{\bh}_k = [\hat{\bh}_{1k}\trans \cdots \hat{\bh}_{Lk}\trans]\trans$. Accordingly, it holds $\tilde{\bh}_{k}\! =\! \bh_{k}\! -\! \hat{\bh}_{k}\! \sim\! \CN(\bzero,\bC_k)$, with $\bC_k = \text{diag}(\bC_{1k}, \ldots, \bC_{Lk})$. 

\medskip

\textbf{User-Centric Uplink Data Transmission.}
For uplink data decoding, we follow a user-centric approach, where each user is  served by a limited number of APs selected among those with the largest large-scale fading coefficient. Such a user-centric association can be mathematically handled by a set of auxiliary diagonal matrices $\bD_{lk} \in \C^{M \times M}, \forall l, \forall k$. Specifically, $\bD_{lk} = \bI_M$, if AP $l$ serves user $k$, and $\bD_{lk} = \bzero_M$, if AP $l$ does not serve user $k$. 
Letting $\bD_k = \text{diag}(\bD_{1k}, \ldots, \bD_{Lk})$ be a block-diagonal matrix, $p_k$ be the uplink transmit power of user $k$, and $\sigma^2 \bI_{LM}$ be the covariance matrix of the collective noise vector,     
 an achievable uplink SE for user $k$, assuming centralized coherent processing at the CPU, is expressed as \cite[Section 5.1]{cellfreebook}:
\begin{equation} \label{eq:SE}
\overline{\mathsf{SE}}_k = \frac{\tu}{\tc} \EX{\log_2 (1 + \mathsf{SINR}_k)}, \quad \text{bit/s/Hz}
\end{equation} 
where
\begin{equation} \label{eq:SINR}
\mathsf{SINR}_k = \frac{p_k |\bv_{k}\herm \bD_{k} \hat{\bh}_k|^2}{\sum\nolimits_{i \neq k}^K p_i |\bv_{k}\herm \bD_{k} \hat{\bh}_i|^2\! +\! \bv_{k}\herm \bZ_k \bv_{k}\! +\! \sigma^2 \norm{\bD_k\bv_{k}}^2}
\end{equation}
with $\bZ_k = \sum\nolimits_{i=1}^K p_i  \bD_k \bC_i \bD_k.$
As to the $k$-th user combining vector $\bv_{k} \in \C^{LM}$, we consider here the so-called Partial MMSE (P-MMSE) combining~\cite[Section 5.1]{cellfreebook}, which suppresses only the most significant user interference contributions. In particular, for an arbitrary user $k$, P-MMSE considers only the interference contributions caused by the users with indices in the set $\mathcal{S}_k \!=\! \{i: \bD_k \bD_i \!\neq\! \bzero_{LM} \}$, and $\bv_k$ is given by    
\begin{equation} \label{eq:PMMSE-combining-vector}
\bv_k \!=\! p_k \left( \sum\limits_{i \in \mathcal{S}_k} p_i \bD_k \hat{\bh}_i \hat{\bh}_i\herm \bD_k \!+\! \bZ_{\mathcal{S}_k}\! +\! \sigma^2 \bI_{LM} \right)^{\!\!\!-1} \!\!\bD_k \hat{\bh}_k,
\end{equation}
where $ \bZ_{\mathcal{S}_k} = \sum\nolimits_{i \in \mathcal{S}_k} p_i  \bD_k \bC_i \bD_k.$

\section{Computation-Offloading and Latency Model}
We assume that any user may receive remote computational support by all the APs of its user-centric cluster and the cloud CPU.
User $k$ needs to execute $w_k$ CPU cycles over $b_k$ computation bits, with a maximum tolerable latency $\mathcal{L}_k$. AP $l$ has a computational capability of $f^{\mathsf{AP}}_l$ CPU cycles per second (\textit{computational rate}). The cloud CPU can execute up to $f^{\mathsf{CPU}}$ instructions per second. The fractions of computational resources assigned to user $k$ by its serving AP $l$ and the CPU are denoted by $f^{\mathsf{AP}}_{l,k}$ and $f^{\mathsf{CPU}}_k$, respectively.
Hence, it holds
\begin{align}
\sum\nolimits_{k=1}^K f^{\mathsf{CPU}}_k &\leq f^{\mathsf{CPU}}, \qquad
\sum\nolimits_{k \in \mathcal{K}_l} f^{\mathsf{AP}}_{l,k} &\leq f^{\mathsf{AP}}_l, \forall l,
\end{align}     
where $\mathcal{K}_l$ is the set including the indices of the users served by AP $l$. 
Let $f_k$ be the overall computational resources assigned to user $k$, given by $f_k = f^{\mathsf{CPU}}_k + \sum_{l \in \mathcal{M}_k} f^{\mathsf{AP}}_{l,k}$, with $\mathcal{M}_k$ being the set including the indices of the APs serving user $k$. Then, $w_k/f_k$ represents the computational time needed to execute $w_k$ CPU cycles at the APs and the cloud CPU (\textit{computational latency}). While, the amount $b_k/R_k$ is the time needed for user $k$ to transmit $b_k$ bits to the APs (\textit{transmission latency}), over the wireless channel  supporting a rate $R_k \! = \! B \times \mathsf{SE}_k$, with $B$ being the transmission bandwidth and $\mathsf{SE}_k$ being the instantaneous SE (i.e., the value attained by~\eqref{eq:SE} with no expectation). Lastly, an additional latency contribution (\textit{fronthaul latency}) is due to the forwarding of the $b_k$ bits from all the APs in the set $\mathcal{M}_k$ to the cloud CPU, over the fronthaul network, which, assuming synchronous transmission across the APs, amounts to $2 b_k M \xi/C_\mathsf{FH}$, where $\xi$ denotes the number of bits used to quantize both real and imaginary parts of the uplink data signal, and $C_\mathsf{FH}$ is the fronthaul capacity of the link between any AP and the cloud CPU, expressed in bit/s.    
Hence, the computational offloading must fulfill the following latency constraint~\cite{Pradhan2020}
\begin{equation} \label{eq:latency-constraint}
\frac{b_k}{R_k} + \frac{w_k}{f_k} + \frac{2 b_k M \xi}{C_\mathsf{FH}} \leq \mathcal{L}_k, \forall k,
\end{equation}  
where we assume that $\mathcal{L}_k$ includes any delay related to the signalling between AP and cloud CPU, and the time needed to send the computational output back to the user.

\bigskip
\noindent \textit{Joint Radio and Computational Resource Allocation}
\medskip

We jointly optimize the users' transmit powers and the allocated computational resources, minimizing the total uplink power while maximizing the minimum SE throughout the network. This optimization problem can be formulated as:
\begin{subequations} \label{prob:P1}
\begin{align}	
  \mathop {\text{minimize}}\limits_{\substack{\{p_k \},~\nu \\ \{f^{\mathsf{CPU}}_k \},~\{f^{\mathsf{AP}}_{l,k} \}}} & \quad \sum\nolimits_{k=1}^K p_k - \nu \varpi(K)  \label{prob:P1:obj} \\[-1ex]
  \text{s.t.} &\quad \frac{b_k}{B~\mathsf{SE}_k} + \frac{w_k}{f^{\mathsf{CPU}}_k + \sum\limits_{l \in \mathcal{M}_k} f^{\mathsf{AP}}_{l,k}} \leq \mathcal{\widetilde{L}}_k, \forall k, \label{prob:P1:C1} \\[-.5ex]
  			  &\quad \mathsf{SE}_k \geq \nu,~\forall k,\label{prob:P1:C2} \\[-.5ex]
  			  &\quad \sum\nolimits_{k=1}^K f^{\mathsf{CPU}}_k \leq f^{\mathsf{CPU}}, \label{prob:P1:C3} \\[-.5ex]
  			  &\quad \sum\nolimits_{k \in \mathcal{K}_l} f^{\mathsf{AP}}_{l,k} \leq f^{\mathsf{AP}}_l, \forall l, \label{prob:P1:C4} \\[-.5ex]
  			  &\quad f^{\mathsf{CPU}}_k \geq 0, \forall k, \label{prob:P1:C5} \\[-.5ex]
  			  &\quad f^{\mathsf{AP}}_{l,k} \geq 0, \forall k, \forall l, \label{prob:P1:C6} \\[-.5ex]
  			  &\quad 0 \leq p_k \leq p_{\text{max}}, \forall k, \label{prob:P1:C7}
\end{align}
\end{subequations}
where $\mathcal{\widetilde{L}}_k = \mathcal{L}_k - (2 b_k M \xi / C_\mathsf{FH})$, $p_{\text{max}}$ is the maximum transmit power per user, $\nu$ is a new variable which represents the minimum instantaneous SE among the users that has to be maximized, and {\color{black}$\varpi(K)$ is a predetermined coefficient, possibly scaling with $K$, that properly weights the two metrics in the objective function}.
This problem is clearly non-convex with respect to $\{p_k \}$ due to the non-convexity (non-concavity) of the latency constraint~\eqref{prob:P1:C1} and the minimum SE constraint~\eqref{prob:P1:C2}. 
In this regard, we propose a \textit{sequential convex programming}, that is an iterative optimization framework wherein in each iteration we optimize a related approximate convex problem.
The instantaneous (net) SE for user $k$ can be expressed as
\begin{align} \label{eq:SE:concavity}
\mathsf{SE}_k &= \log_2 \left(1+\frac{\mathsf{num}_k(\{p_k\})}{\mathsf{den}_k(\{p_k\})}\right) \nonumber \\
&= \log_2(\mathsf{num}_k(\{p_k\})\!+\!\mathsf{den}_k(\{p_k\}))\!-\!\log_2 (\mathsf{den}_k(\{p_k\})) ,
\end{align}    
where $\mathsf{num}_k(\{p_k\})$ and $\mathsf{den}_k(\{p_k\})$ describe the numerator and the denominator of~\eqref{eq:SINR}. We stress that $\mathsf{SE}_k$ is non-concave with respect to the uplink powers $\{p_k\}$. To overcome this non-concavity, we derive a lower bound for the per-user SE. Let us stack the uplink powers as $\bp = [p_1 \cdots p_K]\trans$.
Since $\log_2(\cdot)$ is increasing and the summation preserves concavity, the SE expression in~\eqref{eq:SE:concavity} is the difference of two concave functions. Recalling that any concave function is upper-bounded by its Taylor expansion around any given point $\bp^{(0)}$, a concave lower-bound of $\mathsf{SE}_k$ is obtained as 
\begin{align} \label{eq:SE-lower-bound}
\mathsf{SE}_k(\bp) &= \widetilde{\mathsf{SE}}_k(\bp,\bp^{(0)}) \nonumber \\
&\geq \log_2 (\mathsf{num}_k(\bp)+\mathsf{den}_k(\bp)) - \log_2 (\mathsf{den}_k(\bp^{(0)})) \nonumber \\
&\qquad - \nabla\trans_{\bp} \log_2(\mathsf{den}_k(\bp))\Big\rvert_{\bp = \bp^{(0)}} (\bp - \bp^{(0)}).
\end{align}
Hence,~\eqref{prob:P1:C1} and~\eqref{prob:P1:C2} are approximated and convexified by
\begin{align}
\frac{b_k}{B~\widetilde{\mathsf{SE}}_k(\bp,\bp^{(0)})} + \frac{w_k}{f_k} \leq \mathcal{\widetilde{L}}_k, 
\; \mbox{and} \; \,
\widetilde{\mathsf{SE}}_k(\bp,\bp^{(0)}) \geq \nu, 
\end{align}  
for any $k$ and for any feasible choice of $\bp^{(0)}$.
Note that, the arguments above hold if the elements of the receive combining vector do not depend on the uplink transmit powers. However, in general, the receive combining vector can include a further dependency on $\bp$, for instance in the case of P-MMSE receive combining, as shown in~\eqref{eq:PMMSE-combining-vector}. 
In this case, the approximate SE turns out to be a non-linear function of the uplink transmit powers.  
This issue can be simply tackled by treating the combining vectors at the $n$-th iteration of the SCA optimization framework as constant with respect to the current uplink transmit powers $\bp^{(n)}$, and being exclusively function of $\bp^{(n-1)}$.
Let us introduce the vector of the computational rates $\bbf \in \mathbb{Z}^{K+KL}$ defined as
$$\bbf \! =\! [f^{\mathsf{CPU}}_1,\cdots\!,f^{\mathsf{CPU}}_K,f^{\mathsf{AP}}_{1,1}, f^{\mathsf{AP}}_{2,1},\cdots\!,f^{\mathsf{AP}}_{L,1},f^{\mathsf{AP}}_{1,2},f^{\mathsf{AP}}_{2,2},\cdots\!,f^{\mathsf{AP}}_{L,K}]\trans,$$ such that we can write $f_k = \bb\trans_k \bbf, \, \forall k,$ with $\bb_k \in \{0,1\}^{K+KL}$ being a fixed binary vector determined by the user-centric setup. Specifically,
$\bb_k = [\, \be\trans_k \quad \hat{\bb}\trans_k \,]\trans$, where $\be_k$ is the $k$-th column of $\bI_K$ and $\hat{\bb}_k \in \{0,1\}^{KL}$ such that its element, $\hat{\bb}_k((k-1)L+l)$ is either 1 if AP $l$ serves user $k$ with all its antennas or 0 otherwise. The vector $\hat{\bb}_k$ can be easily constructed upon $\bD_{lk}$ as $\hat{\bb}_k((k-1)L+l) = \tr(\bD_{lk})/M, \forall k, \forall l$. Similarly, we define $\bc_l = [\, \bzero\trans_K \quad \hat{\bc}\trans_l \,]\trans$, and construct the vector $\hat{\bc}_l \in \{0,1\}^{KL}$ upon $\bD_{lk}$ as $\hat{\bc}_l((k-1)L+l) = \tr(\bD_{lk})/M, \forall k, \forall l$.
Hence, the optimization problem at the $n$-th iteration of the proposed SCA method can be formulated as
\begin{subequations} \label{prob:P2}
\begin{align}	
  &\!\mathop {\text{min}}\limits_{\substack{\bp^{(n)},~\nu \\ \bbf^{(n)} \in \mathbb{R}_{>0}^{K\!+\!KL}}}  \quad \boldsymbol{1}\trans_K \bp^{(n)} - \nu \varpi(K)  \label{prob:P2:obj} \\[-1ex]
  &\text{s.t.} \quad \frac{b_k}{B\,\widetilde{\mathsf{SE}}_k(\bp^{(n)},\bp^{(n\!-\!1)})\big\vert_{\bv_k^{(n)}(\bp^{(n\!-\!1)})}}\!+\!\frac{w_k}{\bb\trans_k \bbf^{(n)}} \leq \mathcal{\widetilde{L}}_k, \forall k, \label{prob:P2:C1} \\[-.5ex]
  			  &\qquad \widetilde{\mathsf{SE}}_k(\bp^{(n)},\bp^{(n-1)})\big\vert_{\bv_k^{(n)}(\bp^{(n-1)})} \geq \nu,~\forall k,\label{prob:P2:C2} \\[-.5ex]
  			  &\qquad \sum\nolimits_{k=1}^K \bbf^{(n)}(k) \leq f^{\mathsf{CPU}}, \label{prob:P2:C3} \\[-.5ex]
  			  &\qquad \bc_l\trans \bbf^{(n)} \leq f^{\mathsf{AP}}_l, \forall l, \label{prob:P2:C4} \\[-.5ex]
  			  &\qquad \bzero_K \preceq \bp^{(n)} \preceq p_{\text{max}}\cdot \boldsymbol{1}_K, \label{prob:P2:C5}
\end{align}
\end{subequations}
where we use the continuous relaxation of the original constraint $\bbf^{(n)} \in \mathbb{Z}_{>0}^{K+KL}$ (the real-valued entries of the optimal $\bbf$ are then rounded).  
For an arbitrary iteration $n$ of the SCA method, $\widetilde{\mathsf{SE}}_k(\bp^{(n)}, \bp^{(n-1)})$ is a suitable convex approximation of $\mathsf{SE}_k(\bp^{(n)})$, as  
the following properties are fulfilled~\cite{Marks1978}:
\begin{subequations}
\begin{align}
\mathsf{SE}_k(\bp^{(n)}) &\geq \widetilde{\mathsf{SE}}_k(\bp^{(n)}, \bp^{(n-1)}), \quad \forall n, k, \label{eq:sca:prop1} \\ 
\mathsf{SE}_k(\bp^{(n-1)}) &= \widetilde{\mathsf{SE}}_k(\bp^{(n-1)}, \bp^{(n-1)}), \quad \forall n, k, \label{eq:sca:prop2} \\ 
\nabla_{\bp} \mathsf{SE}_k(\bp^{(n-1)}) &= \nabla_{\bp} \widetilde{\mathsf{SE}}_k(\bp^{(n-1)}, \bp^{(n-1)}), \quad \forall n, k, \label{eq:sca:prop3} 
\end{align} 
\end{subequations}
where $\bp^{(n\!-\!1)}$ is the optimal solution to problem~\eqref{prob:P2} at the iteration $n\!-\!1$. According to \cite{Marks1978}, by virtue of the properties~\eqref{eq:sca:prop1},~\eqref{eq:sca:prop2}, the sequence consisting of the objective functions~\eqref{prob:P2:obj} evaluated at the optimal points is monotonically decreasing and converges to a finite limit. Moreover, this limit, due to the property~\eqref{eq:sca:prop3}, is the value that the original objective function~\eqref{prob:P1:obj} attains at a point satisfying the Karush-Kuhn-Tucher (KKT) conditions of problem~\eqref{prob:P1}. 
Due to space limitations, the proof is available
in the journal version of this work~\cite{Interdonato2021b}.
In this work, by properly selecting the simulation settings, we enforce problem~\eqref{prob:P1} to admit a non-empty feasible set satisfying all the constraints~\eqref{prob:P1:C1}-\eqref{prob:P1:C7}. Two approaches for priory assessing the problem feasibility are presented in~\cite{Interdonato2021b}.    
  
\section{Cell-free versus Cellular Massive MIMO} \label{sec:results}

For benchmarking purposes, we extend the proposed 
joint radio and computational resource allocation framework to a traditional multi-cell massive MIMO system. 
For a  cellular network, we can reasonably assume that only the serving AP offers computational facility to its users. Hence, we can formulate the joint radio and computational resource allocation problem for cellular networks as
\begin{subequations} \label{prob:P1:cellular}
\begin{align}	
  \mathop {\text{minimize}}\limits_{\substack{\{p_k \},~\{f^{\mathsf{BS}}_{l,k} \}, \\\{t_l\}}} & \quad \sum\nolimits_{k=1}^K p_k - \varpi(K) \sum\nolimits_{l=1}^L t_l  \label{prob:P1:cellular:obj} \\[-2ex]
  \text{s.t.} &\quad \frac{b_k}{B~\mathsf{SE}_{lk}} + \frac{w_k}{f^{\mathsf{BS}}_{l,k}} \leq \mathcal{L}^{\text{cell}}_k,~\forall k,\, l \in \mathcal{M}_k, \label{prob:P1:cellular:C1} \\[-.5ex]
  			  &\quad \mathsf{SE}_{lk} \geq t_l,~\forall k,\, l \in \mathcal{M}_k,\label{prob:P1:cellular:C2} \\[-.5ex]
  			  &\quad \sum\nolimits_{k \in \mathcal{K}_l} f^{\mathsf{BS}}_{l,k} \leq f^{\mathsf{BS}}_l,~\forall l, \label{prob:P1:cellular:C3} \\[-.5ex]
  			  &\quad f^{\mathsf{BS}}_{l,k} \geq 0,~\forall k,\,l \in \mathcal{M}_k, \label{prob:P1:cellular:C4} \\[-.5ex]
  			  &\quad 0 \leq p_k \leq p_{\text{max}},~\forall k, \label{prob:P1:cellular:C5}
\end{align}
\end{subequations}
where $\mathsf{SE}_{lk}$ represents the instantaneous uplink SE of user $k$ served by AP $l$ for cellular massive MIMO, which is maximized by using the \textit{Local}-MMSE (L-MMSE) receive combining~\cite[Section 5.3]{cellfreebook}. $\mathcal{L}^{\text{cell}}_k$ denotes the maximum tolerable latency for user $k$ in the cellular setup. This latency value is presumably larger than its cell-free counterpart as it does not include the delay produced by the fronthaul signalling. 
Moreover, in~\eqref{prob:P1:cellular}, $f^{\mathsf{BS}}_l$ indicates the computational capability of AP $l$, expressed in CPU cycles per second. Here, ${\mathsf{BS}}$ stands for \textit{base station}, and this notation is used to emphasize the fact that in a cellular network the APs are actually BS with more antennas and higher computational capability, $f^{\mathsf{BS}}_l > f^{\mathsf{AP}}_l$, than the APs of the cell-free setup. While $f^{\mathsf{BS}}_{l,k}$ denotes the fraction of computational resources assigned to user $k$ by its serving BS $l$.
Lastly, $t_l$ represents the minimum instantaneous SE among the users in cell $l$ that has to be maximized. Hence, the joint optimization in~\eqref{prob:P1:cellular} constitutes a per-cell max-min fairness problem with total transmit power minimization.
Problem~\eqref{prob:P1:cellular} can be convexified via sequential optimization, using the same methodology as for its cell-free counterpart.

\medskip
\noindent \textit{Simulation Setup and Numerical Results} \label{subsec:simulation-setup}
\medskip

We consider a coverage area of 1 km$^2$ served by a total number of antennas $N = LM = 400$. For the cellular massive MIMO setup, $L = 4$ BSs with $M = 100$ antennas each. For the cell-free massive MIMO setup, $L = 100$ APs, with $M = 4$ antennas each. For both the setups, APs/BSs are deployed as a regular grid, and a wrap-around simulation technique is used to remove the edge effects of the nominal coverage~area.    
The systems operate at 2~GHz carrier frequency, over a communication bandwidth of 20 MHz. The receiver noise power is -94 dBm.
The TDD coherence block is $\tc\! =\! 200$ samples long, with $\td=0$ and $\tp\! =\! K/2$.  
$K\! =\! 20$ users  are uniformly distributed at random over the coverage area.
A random realization of users' locations defines a network snapshot, and determines a set of large-scale fading coefficients. These are computed according to the 3GPP Urban Microcell model defined in \cite[Table B.1.2.1-1]{LTE2017}, while the correlation matrices $\{ \bR_{lk} \}$ are generated by using the \textit{local scattering} model~\cite{cellfreebook}. %

For both the setups, we assume $p_{\text{p},k} = p_{\text{max}} = 100$~mW,$~\forall k$, and set the initial choice for the feasible transmit powers of the SCA algorithm as $\bp^{(0)}\! =\! p_{\text{max}}\cdot \boldsymbol{1}_K$. Moreover, {\color{black} $\varpi(K)=1$ is a reasonable choice for these settings, with $\{p_k\}$ values expressed in Watt.} 
Pilot allocation and user-centric cluster formation are performed according to the joint pilot assignment and AP (BS)-user association described in~\cite[Section 5.4]{cellfreebook}. 

We assume $f^{\mathsf{CPU}}\!=\!10^{11}$ cycles/s, while $f_l^{\mathsf{AP}}$ are uniformly distributed random integers from the interval $[10^9,~ 10^{10}]$ cycles/s. The fronthaul capacity is $C_{\mathsf{FH}} = 10$ Gbps, and the number of bits for quantization is set as $\xi = 16$. We select $f_l^{\mathsf{BS}} = \left\lceil \big({\sum\nolimits_{1=l}^{L^{\mathsf{AP}}} f_l^{\mathsf{AP}} + f^{\mathsf{CPU}}}\big)/{L^{\mathsf{BS}}}  \right\rceil,$ where $L^{\mathsf{AP}}$ is the number of APs in the cell-free setup, while $L^{\mathsf{BS}}$ is the number of BSs in the cellular setup, that is 100 and 4, respectively.  This ensures the same number of available computational resources in the coverage area between the two considered setups. Lastly, the latency requirements are equal for all the users in the network, $\mathcal{L}_k = 0.5$ s and $\mathcal{L}_k^{\text{cell}} = 0.7$ s. Common to both setups, the number of bits needed to transfer the users' computational tasks, $\{b_k\}$, are uniformly distributed random integers from the interval $[1, 10]$ Mbits, and the number of CPU cycles necessary to run the task itself is set as a linear function of $b_k$, that is $w_k = \alpha \, b_k$, with $\alpha = 50$ cycles/bit~\cite{Pradhan2020}.

\begin{figure*}[!t]\centering %
	\hspace*{4cm}%
	\csubfloat{\includegraphics[width=.70\columnwidth]{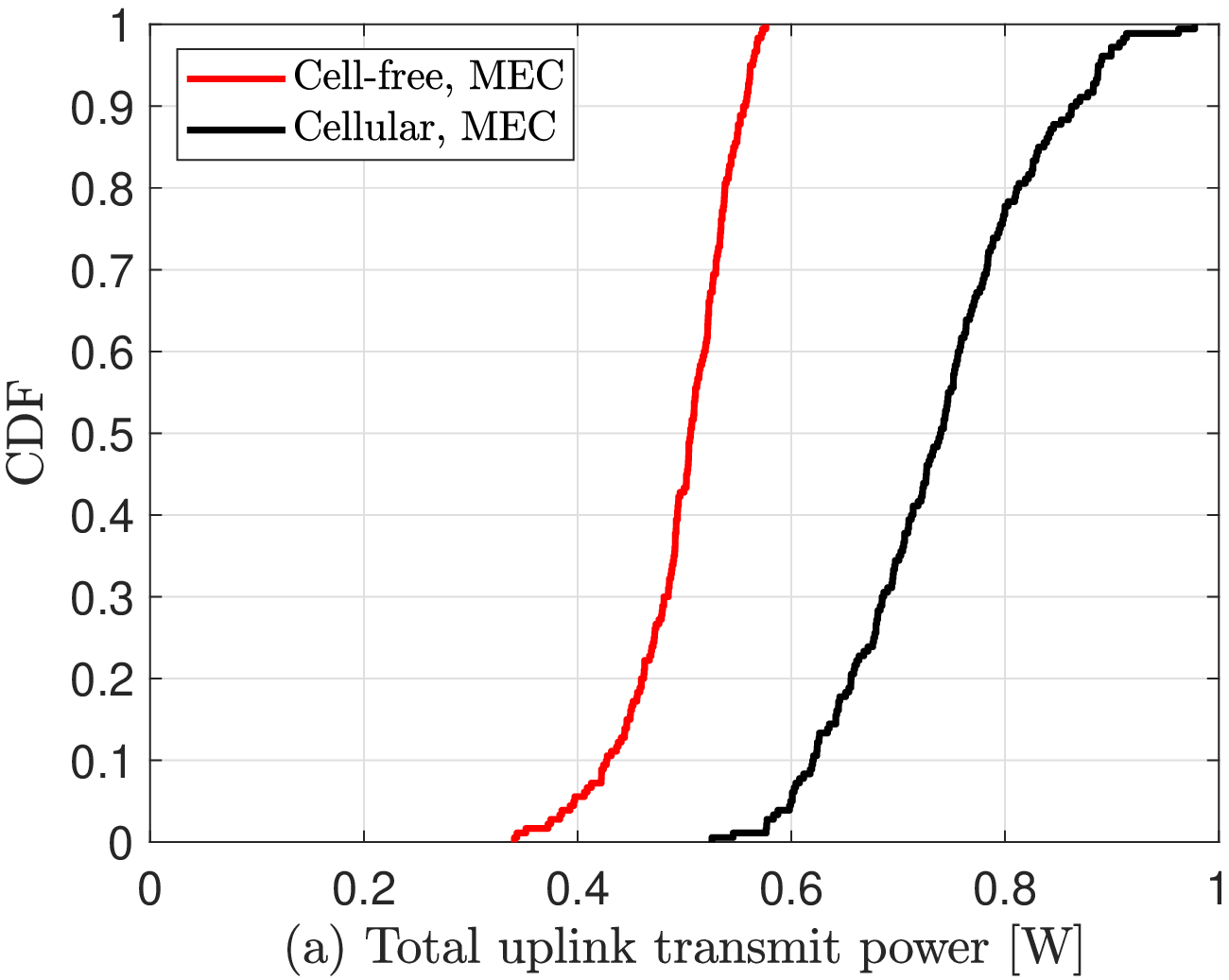}}\centerhfill%
    \csubfloat{\includegraphics[width=.72\columnwidth]{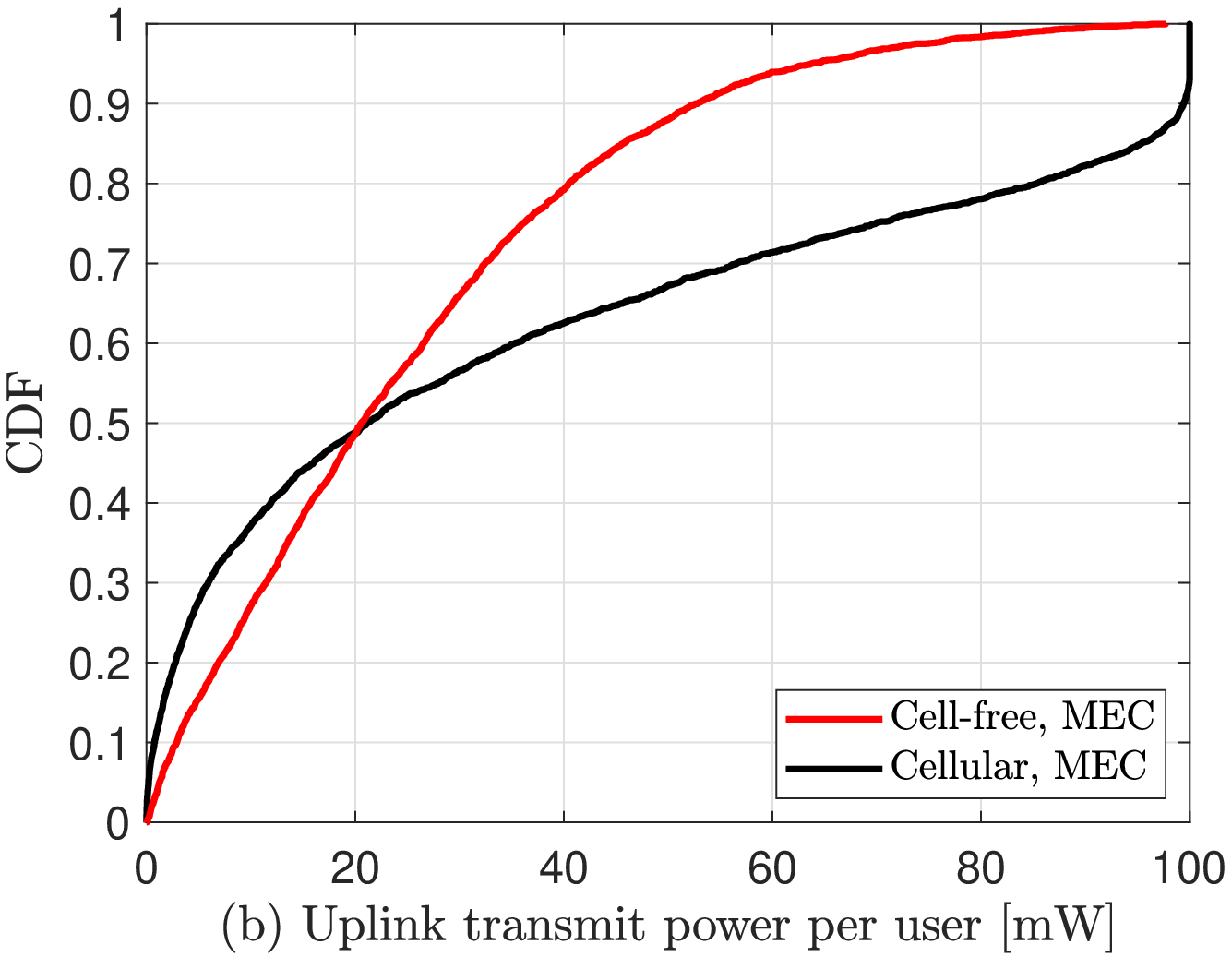}}\centerhfill%
    \\ \vspace*{-2mm}%
	\caption{CDF of the total (a) and per-user (b) uplink transmit power set by the proposed MEC strategy, for cell-free and cellular massive MIMO. The CDFs are obtained over 200 network snapshots. $K=20$, and $p_{\text{max}} = 100$ mW for all the users.}%
	\label{fig:fig1}%
	\vspace*{-5mm}
\end{figure*}

Firstly, we focus on the radiated power consumption. In~\Figref{fig:fig1}(a), we show the cumulative distribution function (CDF), obtained over 200 network snapshots, of the total uplink power, expressed in Watt. This is computed as $\sum_{k=1}^K p_k$, with the uplink powers $\{p_k\}$ being the optimal solutions from the proposed SCA algorithm.      
Numerical results reveal a dramatic transmit power saving for the cell-free massive MIMO users as compared to the cellular massive MIMO users. 
Capitalizing on the distributed topology and the proximity to the user provided by cell-free massive MIMO, our power allocation strategy for MEC offloading enables a total uplink power saving of 30\%--40\% with respect to its centralized counterpart in cellular massive MIMO, and 70\%--80\% with respect an uplink full power transmission strategy (whose total transmit power consumption amounts 2 W). \Figref{fig:fig1}(b) shows the CDF of the uplink transmit power per user, expressed in mW, resulting from the proposed joint power and computational resource allocation.
At high percentiles (i.e., the upper half of the CDF plot), where the transmit power consumption is more significant, we observe that the cell-free massive MIMO users can considerably reduce their transmit power as compared to the cellular massive MIMO users. Interestingly, about 10\% of the cellular users, presumably those at the cell-edge, need to transmit their computational task with full power, whereas the unlucky 10\% of the cell-free users is likely to employ 50--80 mW. The transmit power saving at high percentiles can amazingly reach 50\%.  
On the other hand, the lucky cellular users located at the cell center can leverage the massive BS array gain to substantially reduce their transmit power. This motivates the opposite trend at low percentiles, where the transmit power of the cell-free users is slightly higher. Nevertheless, this negative gap is negligible compared to the performance improvement at high percentiles.
\begin{figure*}[!t] \centering
	\hspace*{4cm}%
	\csubfloat{\includegraphics[width=.72\columnwidth]{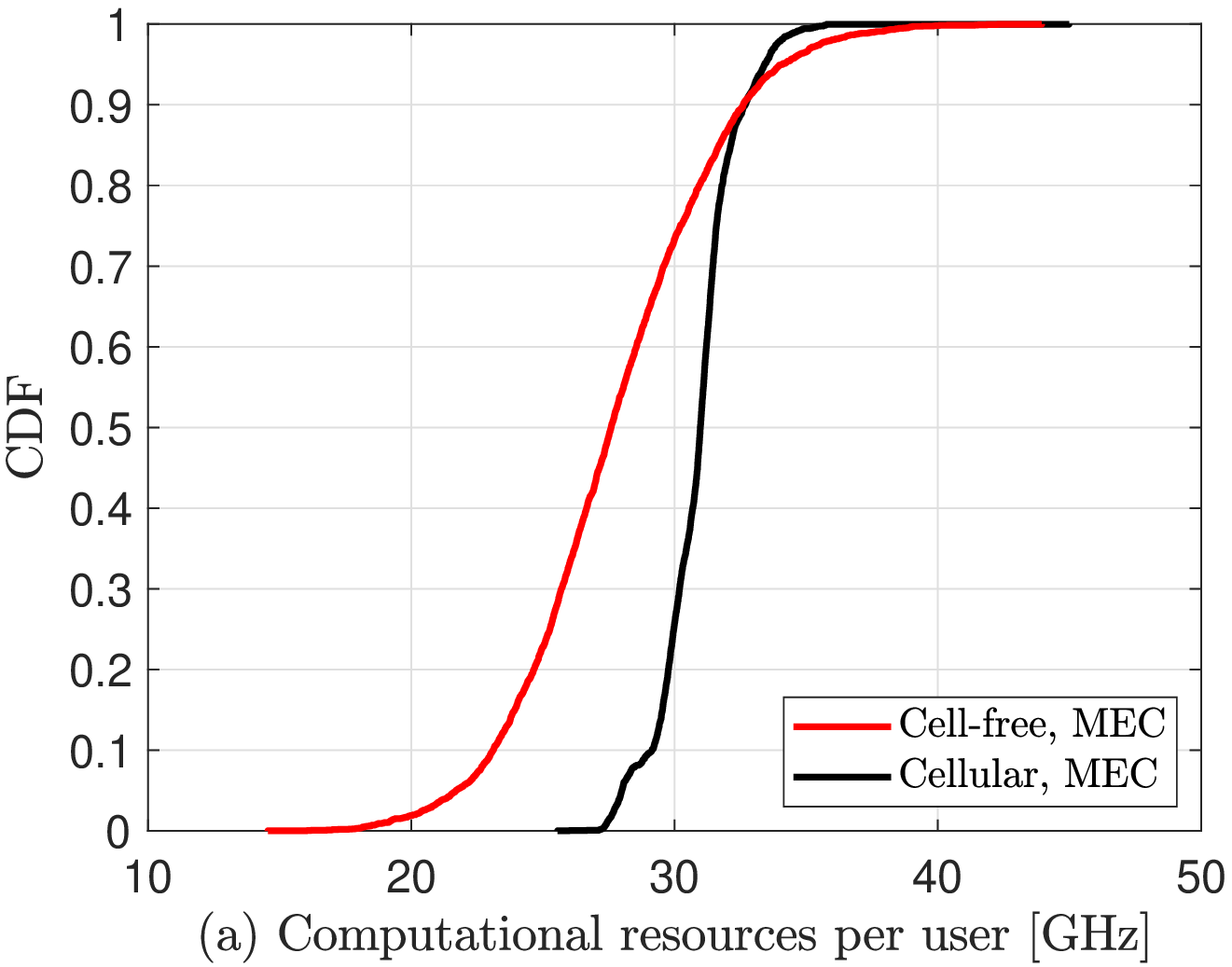}}\centerhfill%
    \csubfloat{\includegraphics[width=.72\columnwidth]{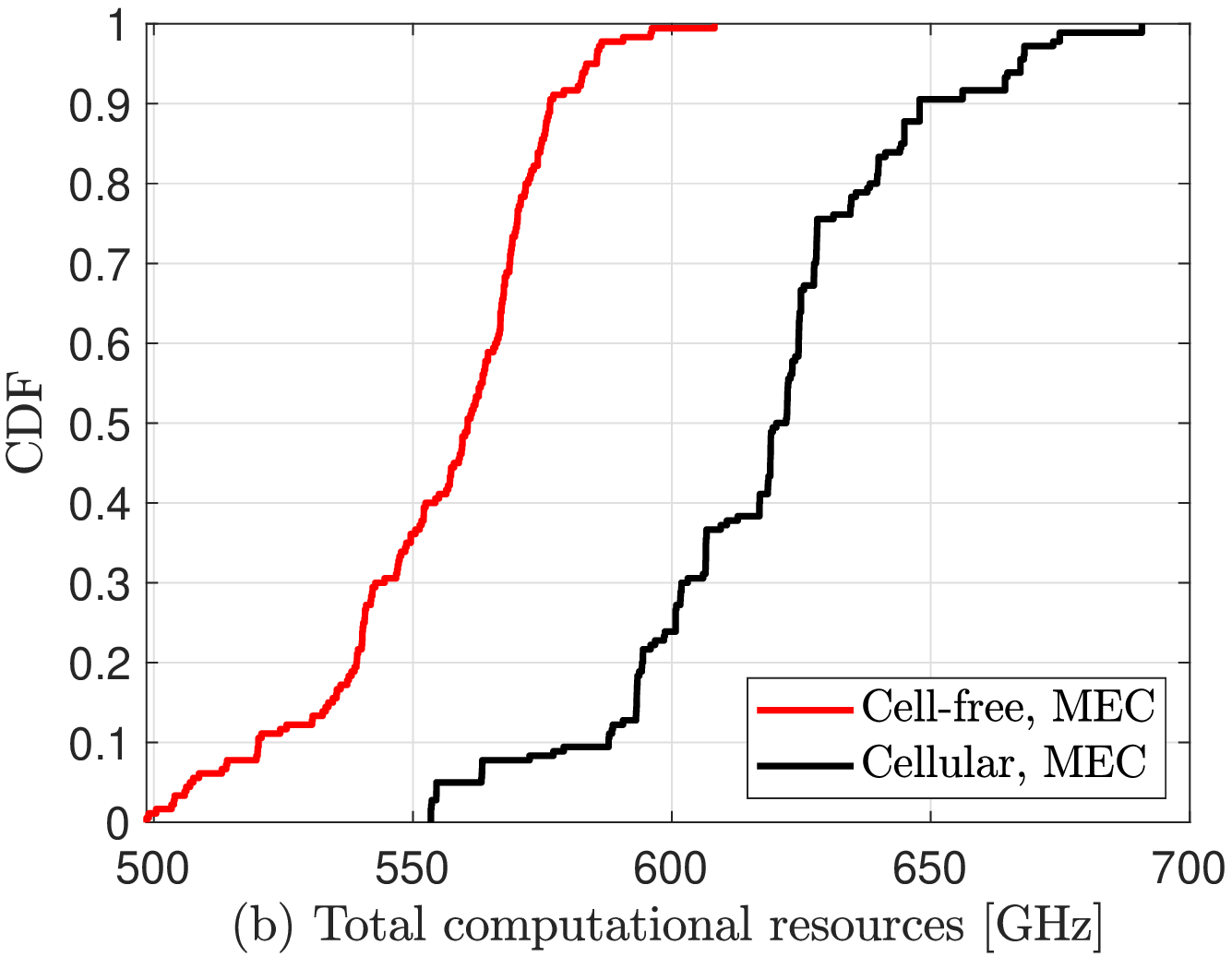}}\centerhfill%
    \\ \vspace*{-2mm}%
	\caption{CDF of the per-user (a) and total (b) computational resources allocated by the proposed MEC strategy, for cell-free and cellular massive MIMO. $f^{\mathsf{CPU}}\!=\!10^{11}$ Hz, $f_l^{\mathsf{AP}}\!\!\sim\mathcal{U}(10^9,\, 10^{10})$ Hz, $f_l^{\mathsf{BS}}\!=\!\left\lceil \big({\sum\nolimits_{1=l}^{L^{\mathsf{AP}}} f_l^{\mathsf{AP}}\!+\!f^{\mathsf{CPU}}}\big)/{L^{\mathsf{BS}}}  \right\rceil$, and $\mathcal{L}_k = 500$ ms and $\mathcal{L}_k^{\text{cell}} = 700$ ms $\forall k.$}
	\label{fig:fig2} %
	\vspace*{-6mm}
\end{figure*} %

\begin{figure*}[!t] \centering
	\hspace*{4cm}%
	\csubfloat{\includegraphics[width=.73\columnwidth]{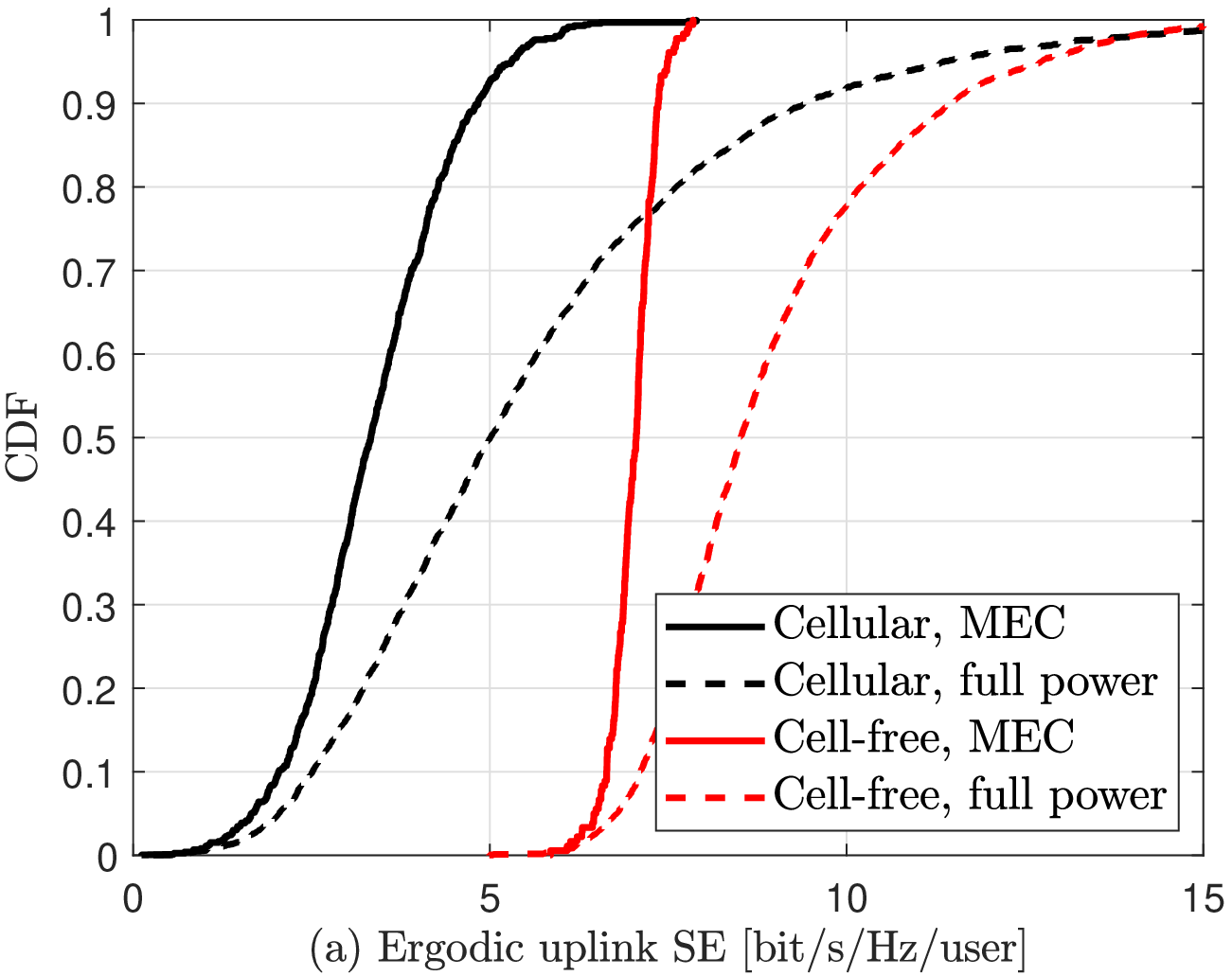}}\centerhfill%
    \csubfloat{\includegraphics[width=.73\columnwidth]{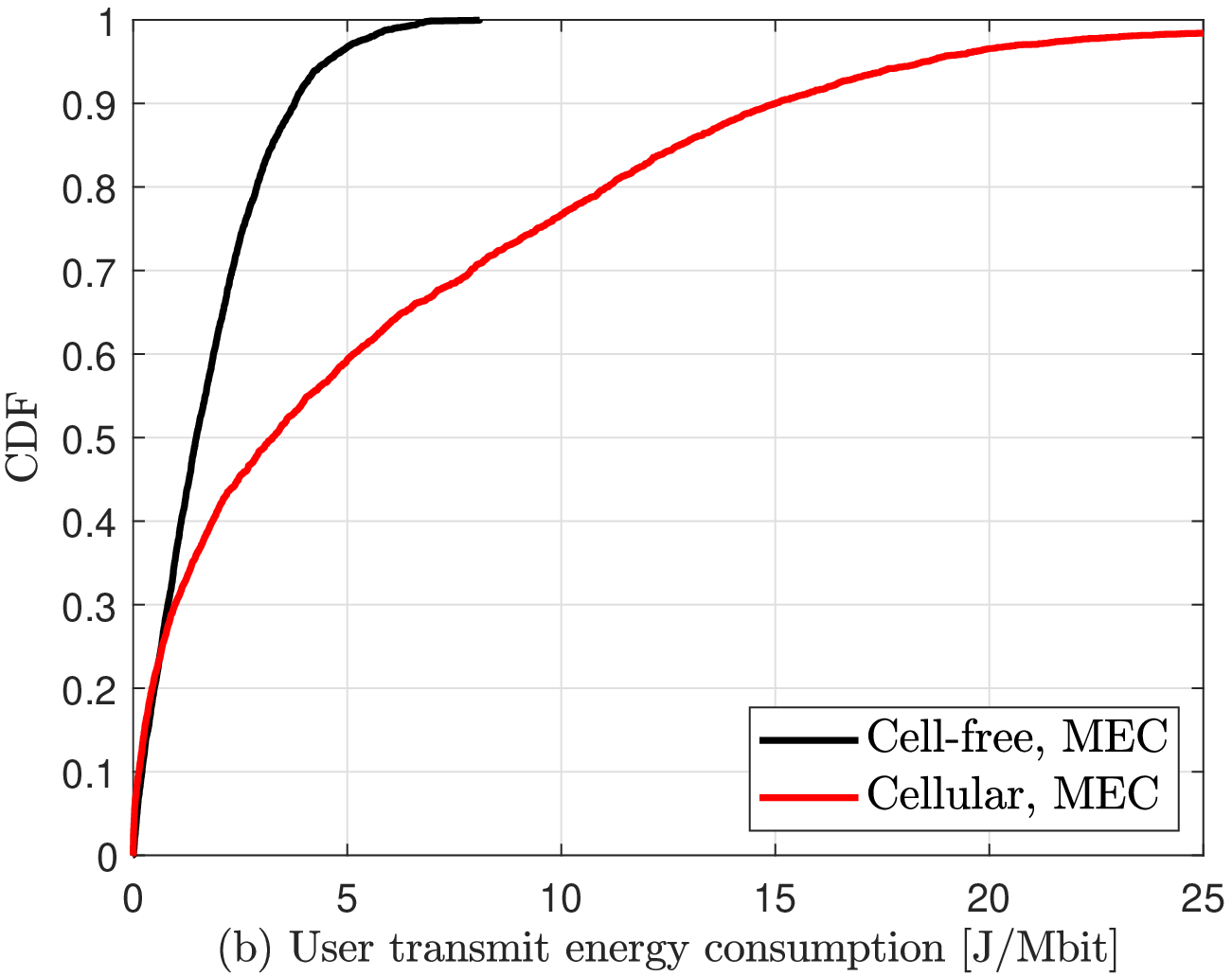}}\centerhfill%
    \\ \vspace*{-2mm}%
	\caption{(a) Ergodic uplink SE per user for cell-free and cellular massive MIMO, either with the proposed MEC offloading strategy or assuming full power transmissions. (b) Transmit energy consumption for cell-free and cellular users.}%
	\label{fig:fig3}%
	\vspace*{-5mm}%
\end{figure*}

Secondly, we focus on how the remote computational resources are allocated.
In \Figref{fig:fig2}(a), we show the CDF of the computational resources allocated to the single user by the proposed MEC offloading strategy, expressed in GHz ($10^9\times$cycles/s). While,~\Figref{fig:fig2}(b) shows the CDF of the total allocated computational resources. These are computed as $\sum_{k=1}^K f^{\mathsf{BS}}_{l,k}$ and $\sum_{k=1}^K (f^{\mathsf{CPU}}_k + \sum_{l \in \mathcal{M}_k} f^{\mathsf{AP}}_{l,k})$ for cellular and cell-free massive MIMO, respectively, and with the computational rates being the optimal solutions of the proposed SCA algorithm.        
User's computational task size being equal, a smaller amount of remote CPU cycles/s is allocated to the cell-free users as compared to that of the cellular users, while fulfilling the latency requirements. This saving is particularly significant at low percentiles, reaching almost 40\%, while is more uniform in terms of total allocated computational resources, ranging from 9\% to 15\%. 
To justify this double gain of the cell-free setup over the cellular setup, we need to focus on the crucial role the SE plays in our formulation. Starting from the latter, in \Figref{fig:fig3}, we show the ergodic uplink SE per user for cell-free and cellular massive MIMO.
Dashed curves represent the SEs obtained by simply assuming uplink full power transmissions for all the users of both the setups. These curves exclusively serve as reference. Additionally, solid curves represent the SEs obtained by adopting the power allocation resulting from the proposed MEC offloading strategy. At first glance, we observe a degradation of the SE with a steeper CDF curve when MEC offloading is performed. This behaviour is a direct consequence of the optimization problem formulation~\eqref{prob:P1} and~\eqref{prob:P1:cellular}. Indeed, minimizing the total uplink power generally leads to a SE degradation, while our max-min fairness approach tends to flatten the CDF of the ergodic SE, as it produces uniform instantaneous SE values among the users. Our choice is motivated by the fact that, when sending the bits of the computational tasks for remote offloading, user's priority is saving as much transmit power as possible to prolong its battery lifetime, while still capitalizing on a fairly good SE to reduce its transmission latency, and more easily fulfill its latency requirements. Hence, our joint power and computational resource allocation provides an optimal balance between uplink transmit power consumption and uplink SE. 
Importantly, considering only the performance of the proposed MEC offloading strategy in \Figref{fig:fig3}(a), the uplink SE of the single cell-free user is, thanks to the macro-diversity gain, remarkably larger than that of the cellular user, for instance 2$	\times$ and 4$\times$ in terms of median and 95\%-likely SE, respectively.
This significant SE gain reflects on the allocation of the computational resources and justifies the numerical results in \Figref{fig:fig2}. Indeed, a larger SE entails a shorter transmission latency (first term of~\eqref{prob:P1:C1}), and gives the opportunity to allocate less computational resources since a longer computational latency (second term of~\eqref{prob:P1:C1}) can be afforded in compliance with the latency constraint. 
The excellent balance between uplink power consumption and SE provided by our joint power and computational resource allocation is clearly shown in \Figref{fig:fig3}(b) which plots the CDF of the (normalized) transmit energy consumption, expressed in J/Mbit, for cell-free and cellular users. For user $k$, this is obtained as
$\mathsf{E}_k\! =\! p_k/(B \,\mathsf{SE}_k)$ (or $\mathsf{SE}_{lk}$ for the cellular setup), with $p_k$ being the optimal solution of the proposed SCA problem, and the SE being computed accordingly. As \Figref{fig:fig3}(b) shows, cell-free users can save a dramatic amount of transmit energy with respect to the cellular users, limiting their energy consumption to 5 J/Mbit. %

\section{Conclusion}
In this work, we addressed the joint optimization of uplink powers and computational resources in a MEC-enabled cell-free massive MIMO system. 
We provided a detailed performance comparison between the proposed MEC-enabled cell-free massive MIMO architecture and its cellular counterpart. Numerical results have revealed the effectiveness of the proposed resource allocation strategy, which enables a significant reduction of users' transmit power and energy consumption, users' transmission latency thanks to higher achievable SEs, and allocated remote computational resources. {\color{black} Future works may include dynamic strategies for computation offloading problems to handle users' mobility.} 

\section*{Acknowledgement}
This paper was supported by the MIUR PRIN 2017 Project ``LiquidEdge''.

\vspace*{-2mm}
\bibliographystyle{IEEEtran}
\bibliography{IEEEabrv,refs-abbr}

\begin{thebibliography}{10}
\providecommand{\url}[1]{#1}
\csname url@samestyle\endcsname
\providecommand{\newblock}{\relax}
\providecommand{\bibinfo}[2]{#2}
\providecommand{\BIBentrySTDinterwordspacing}{\spaceskip=0pt\relax}
\providecommand{\BIBentryALTinterwordstretchfactor}{4}
\providecommand{\BIBentryALTinterwordspacing}{\spaceskip=\fontdimen2\font plus
\BIBentryALTinterwordstretchfactor\fontdimen3\font minus
  \fontdimen4\font\relax}
\providecommand{\BIBforeignlanguage}[2]{{%
\expandafter\ifx\csname l@#1\endcsname\relax
\typeout{** WARNING: IEEEtran.bst: No hyphenation pattern has been}%
\typeout{** loaded for the language `#1'. Using the pattern for}%
\typeout{** the default language instead.}%
\else
\language=\csname l@#1\endcsname
\fi
#2}}
\providecommand{\BIBdecl}{\relax}
\BIBdecl

\bibitem{Ngo2017b}
H.~Q. {Ngo}, A.~{Ashikhmin}, H.~{Yang}, E.~G. {Larsson}, and T.~L. {Marzetta},
  ``Cell-free massive {MIMO} versus small cells,'' \emph{{IEEE} Trans. Wireless
  Commun.}, vol.~16, no.~3, pp. 1834--1850, Mar. 2017.

\bibitem{Interdonato2019}
G.~Interdonato, E.~Bj{\"o}rnson, H.~Q. Ngo, P.~Frenger, and E.~G. Larsson,
  ``Ubiquitous cell-free massive {MIMO} communications,'' \emph{EURASIP J.
  Wireless Commun. and Netw.}, vol. 2019, no.~1, p. 197, 2019.

\bibitem{Buzzi2019c}
S.~{Buzzi}, C.~{D'Andrea}, A.~{Zappone}, and C.~{D'Elia}, ``User-centric {5G}
  cellular networks: {R}esource allocation and comparison with the cell-free
  massive {MIMO} approach,'' \emph{{IEEE} Trans. Wireless Commun.}, vol.~19,
  no.~2, pp. 1250--1264, Feb. 2020.

\bibitem{cellfreebook}
{\"{O}}.~T. Demir, E.~Bj{\"{o}}rnson, and L.~Sanguinetti, ``Foundations of
  user-centric cell-free massive {MIMO},'' \emph{Foundations and
  Trends{\textregistered} in Signal Processing}, vol.~14, no. 3-4, pp.
  162--472, 2021.

\bibitem{Yang2008}
K.~Yang, S.~Ou, and H.-H. Chen, ``On effective offloading services for
  resource-constrained mobile devices running heavier mobile internet
  applications,'' \emph{{IEEE} Commun. Mag.}, vol.~46, no.~1, pp. 56--63, Jan.
  2008.

\bibitem{Mach2017}
P.~Mach and Z.~Becvar, ``Mobile edge computing: {A} survey on architecture and
  computation offloading,'' \emph{{IEEE} Commun. Surveys Tuts.}, vol.~19,
  no.~3, pp. 1628--1656, 3rd Quart. 2017.

\bibitem{Mao2017}
Y.~Mao, C.~You, J.~Zhang, K.~Huang, and K.~B. Letaief, ``A survey on mobile
  edge computing: {T}he communication perspective,'' \emph{{IEEE} Commun.
  Surveys Tuts.}, vol.~19, no.~4, pp. 2322--2358, 4th Quart. 2017.

\bibitem{You2017}
C.~You, K.~Huang, H.~Chae, and B.-H. Kim, ``Energy-efficient resource
  allocation for mobile-edge computation offloading,'' \emph{{IEEE} Trans.
  Wireless Commun.}, vol.~16, no.~3, pp. 1397--1411, Mar. 2017.

\bibitem{Zeng2020b}
M.~Zeng, W.~Hao, O.~A. Dobre, and H.~V. Poor, ``Delay minimization for massive
  {MIMO} assisted mobile edge computing,'' \emph{{IEEE} Trans. Veh. Technol.},
  vol.~69, no.~6, pp. 6788--6792, Jun. 2020.

\bibitem{Ren2018}
J.~Ren, G.~Yu, Y.~Cai, and Y.~He, ``Latency optimization for resource
  allocation in mobile-edge computation offloading,'' \emph{{IEEE} Trans.
  Wireless Commun.}, vol.~17, no.~8, pp. 5506--5519, Aug. 2018.

\bibitem{Sardellitti2015}
S.~Sardellitti, G.~Scutari, and S.~Barbarossa, ``Joint optimization of radio
  and computational resources for multicell mobile-edge computing,'' \emph{IEEE
  Trans. Signal Inf. Process. Netw.}, vol.~1, no.~2, pp. 89--103, Jun. 2015.

\bibitem{Nguyen2019}
T.~T. Nguyen, L.~Le, and Q.~Le-Trung, ``Computation offloading in {MIMO} based
  mobile edge computing systems under perfect and imperfect {CSI} estimation,''
  \emph{IEEE Trans. Services Comput.}, vol.~14, no.~6, pp. 2011--2025, Nov.
  2021.

\bibitem{Pradhan2020}
C.~Pradhan, A.~Li, C.~She, Y.~Li, and B.~Vucetic, ``Computation offloading for
  {IoT} in {C-RAN}: {O}ptimization and deep learning,'' \emph{{IEEE} Trans.
  Commun.}, vol.~68, no.~7, pp. 4565--4579, Jul. 2020.

\bibitem{Hao2019}
Y.~Hao, Q.~Ni, H.~Li, and S.~Hou, ``Energy-efficient multi-user mobile-edge
  computation offloading in massive {MIMO} enabled {HetNets},'' in \emph{Proc.
  IEEE Int. Conf. on Commun. (ICC)}, May 2019, pp. 1--6.

\bibitem{Mukherjee2020}
S.~Mukherjee and J.~Lee, ``Edge computing-enabled cell-free massive {MIMO}
  systems,'' \emph{{IEEE} Trans. Wireless Commun.}, vol.~19, no.~4, pp.
  2884--2899, Apr. 2020.

\bibitem{Marks1978}
B.~R. Marks and G.~P. Wright, ``A general inner approximation algorithm for
  nonconvex mathematical programs,'' \emph{Operations Research}, vol.~26,
  no.~4, pp. 681--683, Aug. 1978.

\bibitem{Interdonato2021b}
\BIBentryALTinterwordspacing
G.~Interdonato and S.~Buzzi, ``Joint optimization of uplink power and
  computational resources in mobile edge computing-enabled cell-free massive
  {MIMO},'' \emph{CoRR}, vol. abs/2111.04678, 2021. [Online]. Available:
  \url{https://arxiv.org/abs/2111.04678}
\BIBentrySTDinterwordspacing

\bibitem{LTE2017}
3GPP, \emph{Further advancements for {E-UTRA} physical layer aspects ({R}elease
  9)}.\hskip 1em plus 0.5em minus 0.4em\relax {3GPP} {TS} 36.814, Mar. 2017.

\end{thebibliography}

\end{document}